\documentclass[%
 reprint,
 amsmath,amssymb,
 aps,
 longbibliography,
]{revtex4-1}
\usepackage[sort&compress]{natbib}
\usepackage[setpagesize=false,dvipdfm]{hyperref}
\usepackage{url}
\usepackage{amsmath}
\usepackage[dvips]{graphicx}
\usepackage{fancybox,ascmac}
\usepackage{graphicx}
\usepackage{amssymb}
\usepackage{bm}
\usepackage{color}
\usepackage{graphicx}
\usepackage{dcolumn}
\usepackage{bm}
\usepackage{float}

\begin{document}
\preprint{APS/123-QED}

\title{Analog and digital codes in the brain}

\author{Yasuhiro Mochizuki}
\email{yasuhiro.mochizuki87@gmail.com}
\author{Shigeru Shinomoto}
\email{shinomoto@scphys.kyoto-u.ac.jp}
\affiliation{%
 Department of Physics, Kyoto University, Kyoto 606-8502, Japan
}%

\date{\today}

\begin{abstract}It has long been debated whether information in the brain is coded at the rate of neuronal spiking or at the precise timing of single spikes. Although this issue is essential to the understanding of neural signal processing, it is not easily resolved because the two mechanisms are not mutually exclusive. We suggest revising this coding issue so that one hypothesis is uniquely selected for a given spike train. To this end, we decide whether the spike train is likely to transmit a continuously varying analog signal or switching between active and inactive states. The coding hypothesis is selected by comparing the likelihood estimates yielded by empirical Bayes and hidden Markov models on individual data. The analysis method is applicable to generic event sequences, such as earthquakes, machine noises, human communications, and enhances the gain in decoding signals and infers underlying activities.
\end{abstract}

\maketitle


\section{Introduction}
Sensation and motion are represented and processed in the brain as series of neuronal discharges called firings or spikes~\cite{rieke1999}. In the early 1900s, the number of neuronal discharges in a given time interval was found to be related to the tension in the associated muscle~\cite{adrian1926}. Since then, correlating the rate of neuronal firings with animal behavior has become standard protocol. Other coding hypotheses have also been studied both experimentally~\cite{hubel1962,maunsell1983,richmond1987} and theoretically~\cite{nadal1994,ben1995,bonnasse2008,rubin2010}. Among these alternatives is temporal coding, which emphasizes the importance of precise spike timings~\cite{richmond1987}.

Coding hypotheses have retained researchers' interest, less for unproductive taxonomy purposes, but because they assist our understanding of neuronal information processing in the brain. Theoretically, it has been demonstrated that a discrete (as opposed to continuous) firing rate increases the rate of information transmission, depending on the width of the time window~\cite{smith1971information,lewen2001neural,bethge2003,ikeda2009,mcdonnell2009}. Thus, assuming that neural systems have evolved to maximize their information transmission rate, different areas of the brain may process signals in different ways. The coding problem has become the subject of theoretical modeling. For instance, in attractor network theory, neuronal activity undergoes transitions among quasistationary states~\cite{amit1989,abeles1995,monasson2013}. Attractor states may manifest as distinct changes in the firing condition. Information processing can feasibly be represented by jumping among quasi-stationary states. In particular, the change-point detection of neurons approaches the theoretical optimum~\cite{kim2012}. Of more practical interest, coding identification may lead to improved information decoding, which would benefit real-time applications such as brain machine interfaces.

Nevertheless, rate coding and temporal coding are not clearly delineated because they are not mutually exclusive. For instance, spike timing may be reinterpreted as high firing rate in a small time window. A clearer distinction between the coding types has been suggested, such that any spike train is classifiable into rate or temporal coding, depending on whether the underlying rate varies slowly or rapidly with time, respectively~\cite{theunissen1995, abbott2001}. However, this principle is not directly applicable to data, because the firing rate cannot be uniquely determined from a single spike train. Rather, spiking is generally irregular and sparse, and the underlying rate can be obtained only from multiple spike train analyses in repeated trials. The fine details of original rate fluctuations are easily erased by inter-trial jittering~\cite{brody1999}. For this reason, the coding hypothesis should be identified on a single trial basis.

\begin{figure}[b]
\begin{center}
\includegraphics[width=1\linewidth]{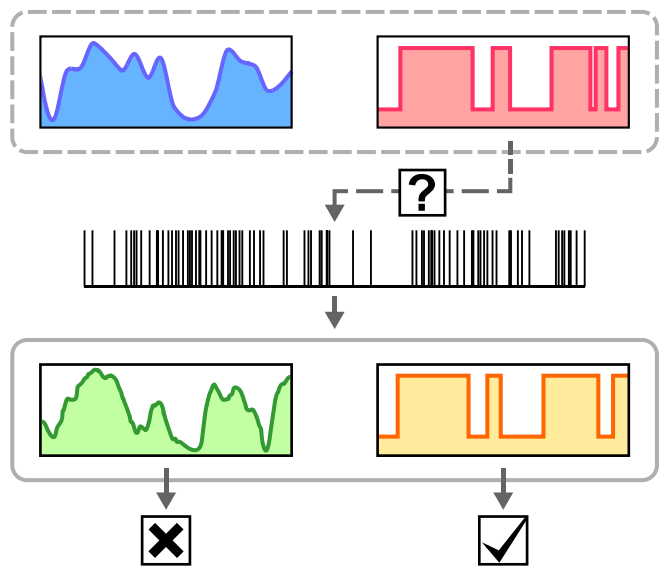}
\caption{ \textbf{Selecting a coding hypothesis for a single spike train}. A spike train is examined to determine whether it is likely transmitting a continuously varying analog signal or discontinuously switching binary signals.}
\label{fig:workflow}
\end{center}
\end{figure}

\begin{figure*}
\begin{center}
\includegraphics[width=1\linewidth]{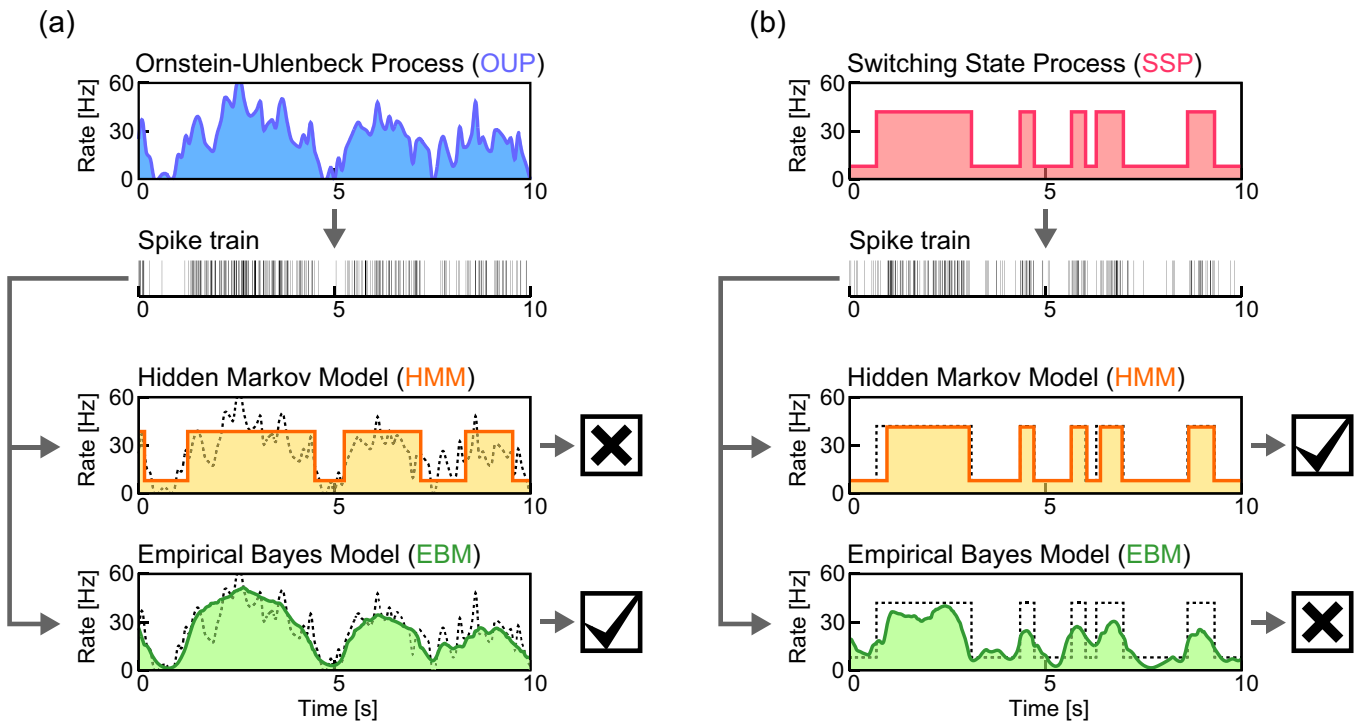}
\caption{\textbf{Continuous and discontinuous rate processes and their decoding.} (a) A spike train is derived from the continuously modulated rate given by the Ornstein--Uhlenbeck process (OUP) (blue). The continuous empirical Bayes model (EBM) (green) estimates the original rate better than the discontinuous hidden Markov model (HMM) (orange). (b) A spike train is derived from the discontinuous rate given by the switching state process (SSP) (red). The HMM better estimates the original rate than the EBM. Simulation parameters are $\mu= 25$ [Hz], $\sigma= 17$ [Hz], and $\tau= 1$ [s] for both the OUP and SSP.}
\label{fig:sample}
\end{center}
\end{figure*}

Here, we suggest a method that selects a unique hypothesis for any given single spike train. The conventional timing-based classification is replaced by an analog--digital classification criterion that inquires whether the spike train is likely transmitting a continuously varying analog signal or discontinuously switching binary signals (Fig.~\ref{fig:workflow}). The proposed classification scheme is similar to the timing-based scheme because analog and digital signals may be represented by the firing rate and timing of spike bursts, respectively. Thus, rather than dispense with the rate-coding hypothesis, we select the best interpretation of a single spike train from alternative rate estimators based on rivalry principles. Specifically, we select a single stochastic model, either the empirical Bayes model (EBM) or the hidden Markov model (HMM), by comparing their likelihood estimates for a given spike train. The EBM and HMM represent the analog and digital codes, respectively. The effectiveness of the inference method is tested on synthetic data derived from inhomogeneous Poisson processes whose continuous and discrete rates are given by the Ornstein--Uhlenbeck process (OUP) and the switching state process (SSP), respectively. Finally, to determine whether different areas of the brain encode signals continuously or discretely, the suggested analysis is applied to biological data.


\section{Modeling spiking processes}
To examine how efficiently the rate estimators infer the underlying rate, we first consider the idealized inhomogeneous Poisson processes, in which spikes are randomly drawn from a given rate function of time. OUP and SSP represent rate functions that fluctuate continuously and discontinuously in time, respectively.

\textbf{The Ornstein--Uhlenbeck process (OUP):} A typical continuously fluctuating process is illustrated in Fig.~\ref{fig:sample}(a). We represent this process by the OUP, originally introduced to describe the fluctuating velocities of Brownian particles. OUP is modeled by the following stochastic differential equation:
\begin{equation}
\frac{1}{2} \frac{d\lambda(t)}{dt}=-\frac{\lambda(t)-\mu}{\tau}+\frac{\sigma}{\sqrt\tau}\xi(t),
\end{equation}
where $\xi(t)$ is the Gaussian white noise characterized by the ensemble average $\langle\xi(t)\rangle=0$ and $\langle \xi(t)\xi(t' )\rangle=\delta(t-t')$. Due to random fluctuations in the OUP, $\lambda(t)$ can be negative even if $\mu$ exceeds the typical fluctuation amplitude $\sigma$. Interpreting $\lambda(t)$ as the temporally fluctuating rate, the firing rate is regarded as zero if $\lambda(t)< 0$.

\textbf{Switching state process (SSP):} A typical discontinuously fluctuating process is illustrated in Fig.~\ref{fig:sample}(b). For this process, we adopt random telegraph state switching, in which the on/off states stochastically alternate under the random telegraph process. Here we consider a symmetric case in which the average inter-transition interval is the same for both states. This process can be realized by repeating the Bernoulli trials for making a transition to another state with the rate $1/\tau$; equivalently, by drawing inter-transition intervals from the exponential distribution with mean $\tau$, $p(t)=\tau^{-1}\exp\left(-t/\tau\right)$. For two states we assign the firing rates $\lambda(t)=\mu-\sigma$ or $\mu+\sigma$ ($\mu >\sigma >0$).

\textbf{1st and 2nd order statistics:} The sample rate processes generated by the OUP and SSP are apparently different (Fig.~\ref{fig:sample}). However, given the same parameter set, $\{\mu,\sigma,\tau\}$, both processes deliver identical first and second order statistics, (mean and correlation function of the rate, respectively), given by

\begin{eqnarray}
\overline{\lambda(t)} &=& \mu, \\
\overline{\delta \lambda(t+s) \delta \lambda(t)} &=& \sigma^2\exp{\left(-\frac{2|s|}{\tau} \right)},
\end{eqnarray}
where the overline represents the time-average and $\delta \lambda(t) \equiv \lambda(t) - \mu$.

\textbf{Inhomogeneous Poisson processes:} Given a time-dependent rate process $\lambda(t)$, a Poisson spike train can be derived by subdividing the time axis into small bins of width $\delta t$ and repeating the Bernoulli trials for generating a spike in every time bin with a probability of $\lambda(t)\delta t$. 

\section{Rate estimators}
Here we introduce EBM and HMM as stochastic models of continuous and discontinuous rate processes, respectively. These models are used as rate estimators.

\textbf{Empirical Bayes model (EBM):} We assume that spikes are independently drawn from the underlying firing rate $\lambda(t)$. In this scenario, termed the inhomogeneous Poisson process, the probability of spike occurrences at times $\{t_j\}\equiv\{t_1,t_2,\dots,t_n\}$ in the period $t\in[0,T]$ is analytically given as~\cite{daleyjones},
\begin{equation}
P(\{t_j \}|\lambda(t))=\left(\prod_{j=1}^n\lambda(t_j)\right)\exp\left(-\int_0^T\lambda(t)dt\right).
\end{equation}
The underlying rate is inferred from the spike train using Bayes' theorem,
\begin{equation}
P(\lambda(t)|\{t_j \})=\frac{P(\{t_j\}|\lambda(t))P(\lambda(t))}{P(\{t_j\})}.
\end{equation}
Here we give a prior distribution functional, by assuming the rate to be generally flat:
\begin{equation}
P_\gamma(\lambda(t))\propto \exp\left(-\frac{1}{2\gamma^2}\int_0^T \left(\frac{d\lambda}{dt}\right)^2dt\right),
\end{equation}
where $\gamma$ is a hyperparameter representing the flatness of the rate process. The hyperparameter can be selected by maximizing the marginal likelihood, or ``evidence,''
\begin{equation}
P_\gamma(\{t_j\}) \equiv \int P(\{t_j\}|\lambda(t))P_\gamma(\lambda(t)) D\{\lambda(t)\},
\label{eq:EBMmarginal}
\end{equation}
where $D\{\lambda(t)\}$ denotes that the functional is integrated over all possible rate processes. If data are provided, this marginalization integral can be maximized with respect to $\gamma$ by the expectation and maximization (EM) algorithm~\cite{smith2003}. The negative of the logarithm of the marginal likelihood corresponds to the free energy~\cite{mackay1992,bruce1994, carlin2010}. With the hyperparameter $\hat{\gamma}$ that maximizes the marginal likelihood or minimizes the free energy, we obtain the maximum a posteriori (MAP) estimate of the rate process, $\hat{\lambda}(t)$, that maximizes the posterior distribution functional, $p_{\hat\gamma}(\lambda(t)|\{t_j\})$~\cite{koyama2010}. An application program for estimating the MAP rate for a given spike train is accessible by Ref.~\cite{shimokawa_app}.

\textbf{Hidden Markov model (HMM):} When estimating the firing rate by HMM~\cite{bishop,shintani2012}, the spike train is derived from the rate of transition between different states according to the Markov process. Here we adopt a two-state HMM model, in which the rate takes one of two values. Model parameters are the two rates $\lambda_1$ and $\lambda_2$, the elements of the transition matrix, and the probabilities of the initial states. These parameters are estimated by the Baum--Welch algorithm, and then the most likely sequence of hidden states is then obtained from the Viterbi algorithm. The estimated rate $\hat{\lambda}(t)$ is regarded as the sequence of alternating rates assigned as the selected hidden states. An application program for performing HMM rate estimation is accessible by Ref.~\cite{mochizuki_app}.

\begin{figure*}
\begin{center}
\includegraphics[width=1\linewidth]{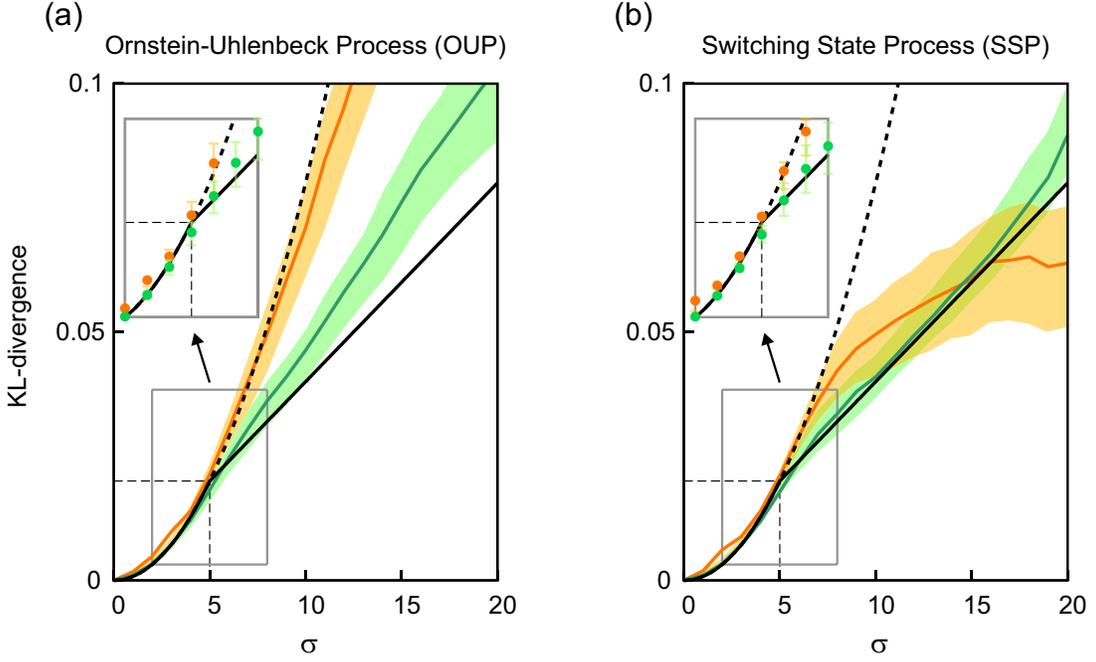}
\caption{\textbf{Kullbac--Leibler (KL) divergence of the estimated distribution $\bm{\hat{p}(t)}$ from the underlying distribution $\bm{p(t)}$.} Numerically estimated EBM and HMM values are depicted in green and orange, respectively. (a) The continuous OUP rate process. (b) The discontinuous SSP rate process. The rates are estimated from spike trains of $n$=1,000 spikes. The black solid line is the analytical result obtained by the path integral, Eq.~(\ref{eq:analyticalEBM}). The edges of the light green and yellow regions represent the upper/lower quartiles of KL divergence estimated from 1,000 samples. Other parameters are $\mu$=25 [Hz], $\tau$=1 [s], giving a theoretical detection limit of $\sigma_c=\sqrt{\mu/\tau}$=5 [Hz].}
\label{fig:kl}
\end{center}
\end{figure*}

\section{Testing the rate estimators using synthetic spike trains}
In this section, we compare the accuracy of EBM and HMM in estimating the underlying rates from OUP- and SSP-derived spike trains, representing continuous and discontinuous rate processes, respectively.

\subsection{The Kullback--Leibler (KL) divergence}
Apparently the firing rate of the spike train derived from OUP is better estimated by EBM than by HMM, while the opposite is true for the SSP-derived spike train (Fig.~\ref{fig:sample}). If the underlying rates of synthetic data are known, the estimation accuracy can be evaluated by directly measuring the deviation of the estimated rate $\hat{\lambda}(t)$ from the underlying rate $\lambda(t)$. Furthermore, if spikes are derived independently from the underlying rate $\lambda(t)$ and the resulting spike train follows a Poisson process, the goodness of the rate estimator is the deviation of the normalized density of individual spikes, $\hat{p}(t) \equiv \hat{\lambda}(t)/\int_0^T \hat{\lambda}(t') dt' $ from the normalized underlying density, $p(t) \equiv \lambda(t)/\int_0^T \lambda(t') dt'$, assuming that the average rate is correctly captured; i.e., $\int_0^T \hat{\lambda}(t') dt'=\int_0^T \lambda(t') dt'$. The deviation of distribution functions may be represented by the Kullback--Leibler (KL) divergence~\cite{cover_elements,mezard2009}, defined as
\begin{equation}
D(p||\hat{p}) \equiv \int_0^T p(t) \log{(p(t)/\hat{p}(t))} dt \ge 0,
\end{equation}
where the equality holds if the two distribution functions are equal. The KL divergence is represented as the surplus of the cross entropy~\cite{murphy_machine},
\begin{equation}
H(p,\hat{p}) \equiv -\int_0^T p(t) \log{\hat{p}(t)} dt,
\label{eq:fe}
\end{equation}
over the entropy of the underlying distribution $p(t)$, 
\begin{equation}
H(p) \equiv -\int_0^T p(t) \log{p(t)} dt,
\end{equation}
that is;
\begin{equation}
D(p||\hat{p}) = H(p,\hat{p}) - H(p).
\end{equation}

\textbf{Rate estimation using the EBM:} In EBM, the KL divergences of the OUP- and SSP-derived spike trains depend similarly on $\sigma$ (the green lines in Figs.~\ref{fig:kl}(a) and (b)). 
The initial quadratic increase with $\sigma$ may be interpreted as follows. If the rate fluctuation in a spike train is small, it will not be detected by any rate estimator. Rather than sampling a fluctuation, principled estimators such as the EBM will draw a fixed rate of $\hat{p}(t)=1/T$, whereby the cross entropy is given as
\begin{equation}
H(p,\hat{p}) = -\int_0^T p(t) \log (1/T) dt = \log T.
\end{equation}
The entropy can be approximated by expanding $p(t)$ in terms of the deviation of the normalized distribution from the mean $1/T$,
\begin{eqnarray}
H(p) &=& - \int_0^T \left( \frac{1+\delta p(t)}{T} \right) \log{\left( \frac{1+\delta p(t)}{T} \right)} dt  \nonumber \\
&\approx& \log T - \overline{\delta p^2}/2 = \log T - \frac{\sigma^2}{2\mu^2}.
\label{eq:ent}
\end{eqnarray}
Thus, in both the OUP and SSP, the KL divergence of the constant probability $\hat{p}=1/T$ is approximated as
\begin{equation}
D_0 = H(p,\hat{p}) - H(p) \approx \frac{\sigma^2}{2\mu^2}.
\end{equation}

\begin{figure}
\begin{center}
\includegraphics[width=1\linewidth]{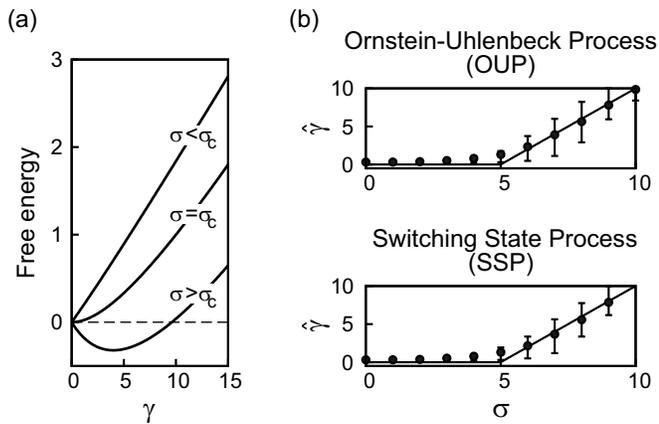}
\caption{\textbf{Detectable-undetectable phase transitions of the empirical Bayes model.} (a) Free energy defined by the log-likelihood Eq.~(\ref{eq:freeenergy_marginal}) for the cases of $\sigma<\sigma_c$, $=\sigma_c$, and $>\sigma_c$. (b) An optimal hyperparameter $\hat{\gamma}$ representing the degree of flatness. The conditions of phase transition for the OUP and SSP are given by Eq.~(\ref{eq:optgamma}). The error bars of numerical data represent the average and upper/lower quartiles of the hyperparameter $\hat{\gamma}$ determined by applying practical optimization algorithms to 1,000 synthetic data numerically generated by simulating the inhomogeneous Poisson processes. Parameters of synthetic data used for numerical analysis are the same as those of Fig.\ref{fig:kl}.}
\label{fig:phase_trans}
\end{center}
\end{figure}

\textbf{Theoretical estimate of the EBM using the path integral:} As the rate fluctuation $\sigma$ increases further, the KL divergence departs downward from this quadratic line, implying that rate fluctuations have begun to be appropriately detected by EBM. The EBM marginalization integral has been shown to be analytically solvable~\cite{koyama2007,koyama2013}. Equation~(\ref{eq:EBMmarginal}) can be transformed into a path integral format as~\cite{bialek1996}
\begin{equation}
P_\gamma(\{t_j\})=\frac{1}{Z(\gamma)}\int D\{\lambda(t)\}e^{-\int_0^T L(\lambda,\dot{\lambda},t)dt},
\end{equation}
where the ``Lagrangian'' $L(\lambda,\dot{\lambda},t)$ is
\begin{equation}
L(\lambda,\dot{\lambda},t)=\frac{1}{2\gamma^2}\dot{\lambda}^2+\lambda-\sum_{i=1}^n \delta(t-t_i)\log\lambda.
\end{equation}

The ``classical path'' corresponding to the MAP estimate of the rate process $\lambda(t)$ is obtained by the Euler--Lagrange equation:
\begin{equation}
\frac{d}{dt} \left(\frac{\partial L}{\partial\dot{\lambda}}\right)-\frac{\partial L}{\partial \lambda}=0.
\end{equation}
Here, the log marginal likelihood is averaged over possible realizations of spike trains derived from a rate function. When deriving a spike train from an underlying rate $\lambda(t)$, the fluctuations in spike counts within a given time interval are Poisson-distributed. In this case, the variance in the spike count equals the mean, and the rate of an individual spike train is described by a stochastic function:
\begin{equation}
\lambda(t)+\sqrt{\lambda(t)}\xi(t),
\end{equation}
where $\xi(t)$ denotes Gaussian white noise. The path integral can be evaluated by expanding the ``action integral'' to quadratic order in the deviation of the rate from the mean~\cite{koyama2007,koyama2013}. The free energy is analytically obtained as
\begin{eqnarray}
F(\gamma) &=& - \left\langle \log P_\gamma(\{t_j\}) \right\rangle \nonumber \\
&\approx& \frac{T|\gamma|}{4\sqrt\mu}\left(1-\frac{2\tau\sigma^2} {2\mu+\gamma\tau\sqrt\mu}\right)+{\rm const.},
\label{eq:freeenergy_marginal}
\end{eqnarray}
where the angle brackets represent the averaging operation with respect to the $\xi(t)$ ensemble. The hyperparameter $\gamma$ may be selected by minimizing the free energy:
\begin{equation}
\hat{\gamma} =\arg\min_{\gamma} F(\gamma)=
\left\{ 
\begin{array}{l l}
    0, & {\rm if }  \,\, \sigma< \sigma_c,\\
    2{(\sigma-\sigma_c)}/{\sqrt{\tau}},  & {\rm otherwise,}
\end{array} \right.
\label{eq:optgamma}
\end{equation}
where $\sigma_c=\sqrt{\mu/\tau}$. Thus, $\gamma$ vanishes, or equivalently, the flatness of the rate diverges, if the fluctuation amplitude of the underlying rate is below the critical value $\sigma_c$, implying that the rate fluctuation is undetectable (Fig.~\ref{fig:phase_trans}).

The MAP estimate of the rate, or solution of the Euler--Lagrange equation, is given as
\begin{equation}
\hat{\lambda}(t) \approx \mu + \frac{\hat{\gamma}}{2\sqrt\mu}\int_0^T (\delta \lambda(s)+\sqrt{\mu}\xi(s))e^{-\frac{\hat{\gamma}|t-s|}{\sqrt\mu}}ds,
\label{maporbit}
\end{equation}
where $\sqrt{\lambda(t)}\xi(s)$ is approximated by $\sqrt{\mu}\xi(s)$. The cross entropy can be obtained by expanding $p(t)$ in terms of the deviation of the normalized distribution, and by averaging over all possible realizations of spike trains $\xi(s)$ as
\begin{eqnarray}
H(p,\hat{p}) &=& - \int_0^T \frac{1+\delta p(t)}{T} \left\langle \log{ \left( \frac{1+\delta \hat{p}(t)}{T} \right)} \right\rangle dt  \nonumber \\
&\approx& \log T - \overline{ \langle \delta p \delta \hat{p} - \delta \hat{p}^2 /2 \rangle }.
\label{freeenergy}
\end{eqnarray}
By calculating the ensemble average, the KL divergence is obtained as
\begin{eqnarray}
D(p||\hat{p}) \approx
\left\{ 
  \begin{array}{l l}
    \sigma^2/(2\mu^2),   & {\rm for} \,\, \sigma< \sigma_c,\\
    \sigma\sigma_c/(2\mu^2),  & {\rm otherwise.}
  \end{array} \right.
\label{eq:analyticalEBM}
\end{eqnarray}
The phase transition at $\sigma=\sigma_c$, above which the fluctuations become detectable, is discernible in the analytical KL divergence curves (the black solid lines in Fig.~\ref{fig:kl}). Though this solution assumes that $\sigma \ll \mu$, it reasonably agrees with the numerical solutions (the green lines in Fig.~\ref{fig:kl}) even when $\sigma$ is comparable to $\mu$.

\textbf{Rate estimation using the HMM:} To enable comparison between the rate-detection performances of EBM and HMM, the KL divergence of HMM is also plotted in Fig.~\ref{fig:kl} (the orange lines).

When $\sigma< \sigma_c$, the KL divergences of the HMM are higher than $D_0=\sigma^2/(2 \mu^2)$ for both OUP- and SSP-derived spike trains. From this result, we infer that HMM needlessly tracked the fluctuations of individual data; accordingly, the rate estimation is inferior to that obtained by simply indicating the constant mean rate.


When rate fluctuations are large, $\sigma> \sigma_c$, the KL divergences obtained by HMM differ widely between the OUP and SSP data. For the SSP, the KL divergence is lower than that of EBM, implying that HMM more accurately estimates the underlying transition rate between the two states. Contrariwise, for the OUP data, which realize continuous rate changes, the performance of EBM is always superior to that of HMM.

\begin{figure}[!b]
\begin{center}
\includegraphics[width=1\linewidth]{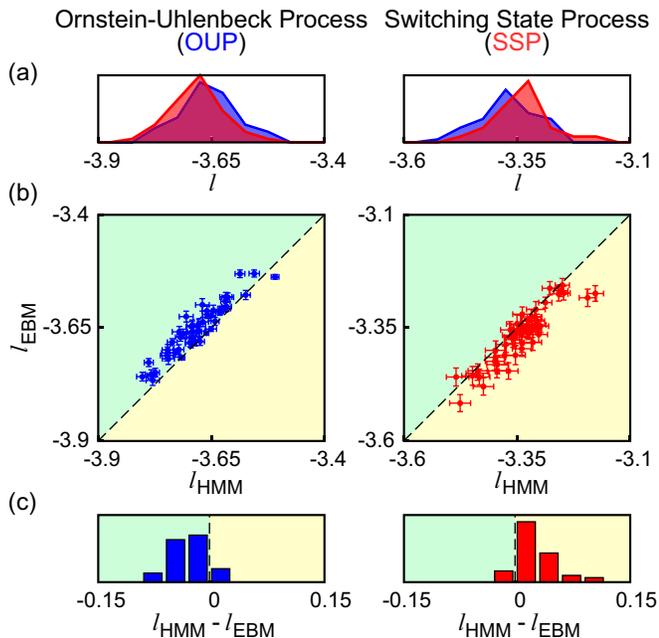}
\caption{\textbf{Validating the likelihoods of analog and digital coding hypotheses.} 
(a) Likelihood distributions of EBM and HMM, (the blue and red distributions, respectively), calculated by Eq.~(\ref{eq:validationlikeli}), for trains of 1,000 spikes derived from continuous OUP and discontinuous SSP. (b) Scatter plots of the likelihoods on the ($l_{\rm HMM}$, $l_{\rm EBM}$)-plane (c) Distribution of the likelihood difference, $l_{\rm HMM} - l_{\rm EBM}$. Model parameters: ($\mu$, $\sigma$, $\tau$) = (25 [Hz], 10 [Hz], 1 [s]) for the OUP and (25 [Hz], 20 [Hz], 1 [s]) for the SSP. Number of spikes $n$=1,000; subsampling spikes $m$=10; sampling trials $k$=100.}
\label{fig:l1000}
\end{center}
\end{figure}

\begin{figure*}
\begin{center}
\includegraphics[width=1\linewidth]{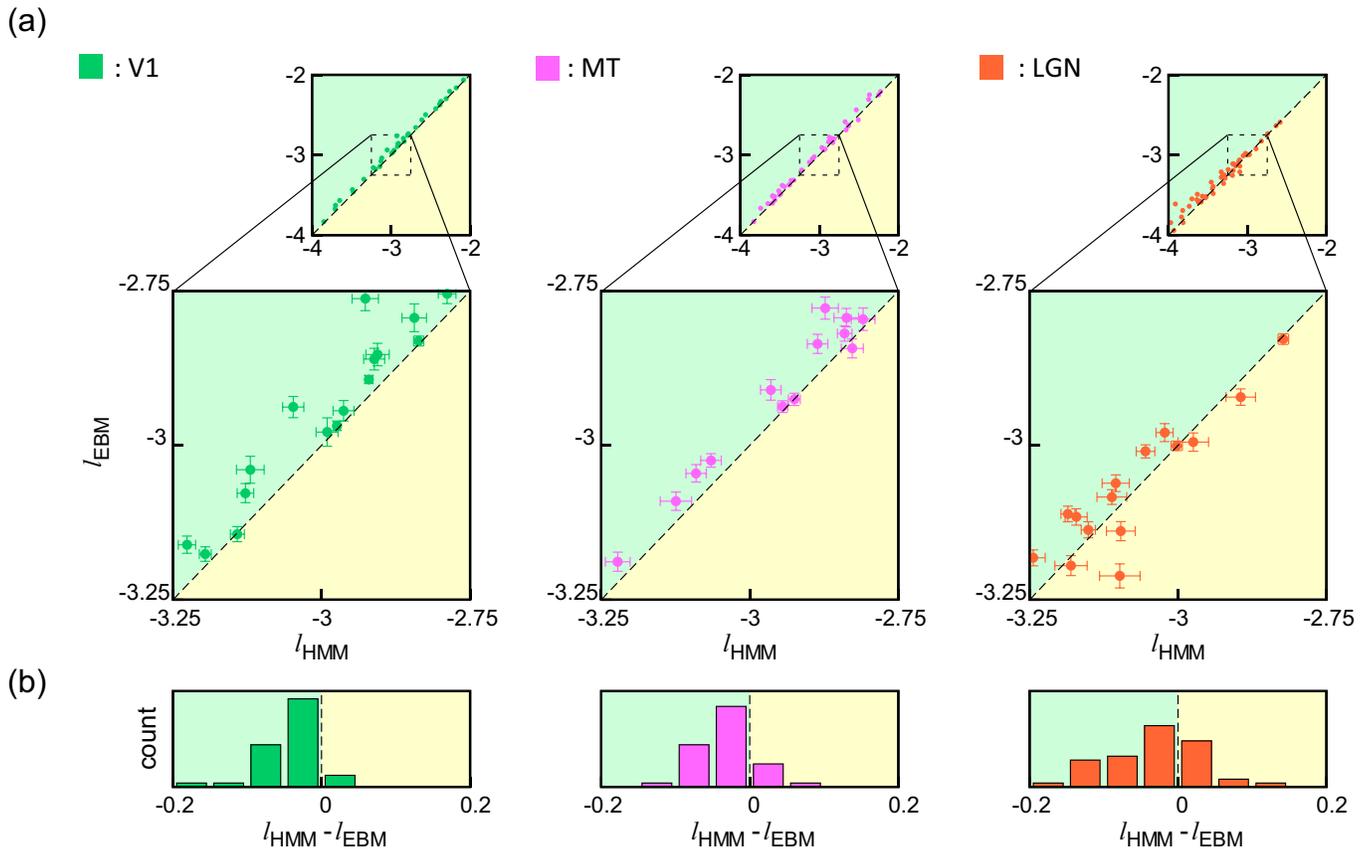}
\caption{\textbf{Analysis of biological neuronal spike trains.} (a): Scatter plots of the log-likelihood of EBM against the log-likelihood of HMM. Each cross represents the standard error of the mean likelihoods obtained for a neuron in V1 (green), MT (pink) and LGN (orange). (b): Histograms of differences between the two log-likelihoods shown in Fig.~\ref{fig:l1000}(c). Number of spikes $n$=1,000; subsampling spikes $m$=10; sampling trials $k$=100.}
\label{fig:likelihoodbio}
\end{center}
\end{figure*}

\begin{figure*}
\begin{center}
\includegraphics[width=1\linewidth]{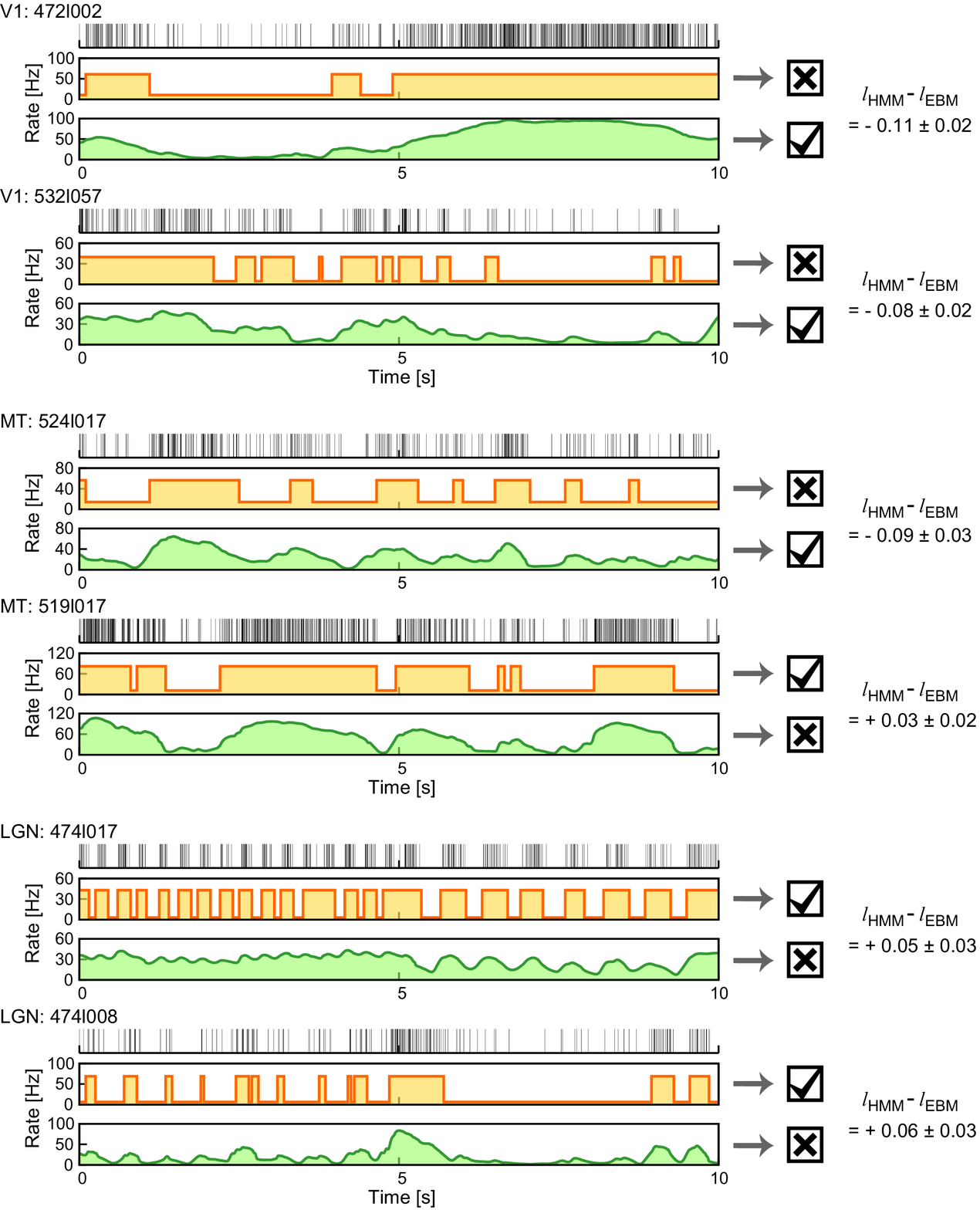}
\caption{\textbf{Sample of biological spike trains and their estimated rates.} Spike trains are recorded from V1, MT, and LGN and the rates are estimated from the continuous and discontinuous estimators, EBM (green) and HMM (orange), respectively. The title of each set of plots indicates the neuron ID in Refs~\cite{movtion_data}, and the rate estimation selected according to the likelihood are shown with the tick marks.}
\label{fig:spiketrainsbio}
\end{center}
\end{figure*}
\subsection{Validating rate estimators}
The KL divergence is an impractical measure because the original rate is not known in real applications. Here we suggest a random subsampling validation method that evaluates the rate estimators in terms of their goodness of estimation.

Given a spike train of $n$ spikes, we randomly remove $m (\ll n)$ spikes and estimate the rate profile $\hat{\lambda}_{n-m}(t)$ from the remaining $n-m$ spikes. Because every spike is independently derived from the underlying rate, the likelihood of the rate profile of the unused $m$ spikes is the product of the likelihoods of normalized densities $\hat{p}(t)=\hat{\lambda}_{n-m}(t)/\int_0^T \hat{\lambda}_{n-m}(t')dt'$. Thus, the log-likelihood per a single spike is estimated as
\begin{equation}
l=\frac{1}{m}\sum_{i=1}^m \log\left(\frac{\hat{\lambda}_{n-m}(t_i)}{\int_0^T \hat{\lambda}_{n-m}(t')dt'} \right).
\label{eq:validationlikeli}
\end{equation}
Repeating this procedure $k$ times, we compute the mean and the standard error of the log-likelihood of a given spike train. The cross-validated log-likelihood should approximate the negative cross entropy, $-H(p, \hat{p})$ (Eq.~(\ref{eq:fe})).

When the distributions of the likelihoods for EBM and HMM are directly compared, their relative superiority or inferiority is not evident (Fig.~\ref{fig:l1000}(a)). This occurs because the entropy $H(p)$ of individual rate processes fluctuates among samples, and the estimated log-likelihood and negative cross entropy $-H(p,\hat{p})$ alone do not reflect the goodness of the rate estimation. Thus, we suggest comparing the likelihoods of EBM and HMM for individual spike trains (Fig.~\ref{fig:l1000}(b)) or computing their difference $l_{\rm HMM} - l_{\rm EBM}$ (Fig.~\ref{fig:l1000}(c)). The conformity of the data to HMM and EBM is now clearly detected even from sequences of O(1,000) spikes, which are typically acquired from experiments.

\section{Analysis of biological data}

Finally, we apply our method to real data. To this end, we analyze biological neuronal spike trains by the cross-validation method. The test is conducted on publicly available spike data. All data were collected from the visual cortical areas, primary visual cortex (V1) and middle temporal area (MT), and from the thalamus and lateral geniculate nucleus (LGN) of monkeys ({\it Macaca fascicularis}) repeatedly presented with a drifting sinusoidal grating~\cite{movtion_paper,movtion_data}. In each trial, the recording times of a single run were 6,000 or 3,000 ms for V1, 1,280 ms for MT, and 5,138 ms for the LGN. Only the trials with mean firing rate greater than 10 Hz were accepted, and spike trains recorded in different trials were concatenated into a final spike train of 1,000 spikes. The numbers of accepted neurons were 39, 40, and 49, respectively for V1, MT, and LGN.

The results of the cross-validation analysis are shown in Fig.~\ref{fig:likelihoodbio}. Fractions of neurons exhibiting analog and digital coding patterns differ between the three brain regions. In particular, more discontinuous firing patterns were observed in LGN neurons than in V1 and MT neurons (15/49 in the LGN, versus 3/39 and 7/40 in the V1 and MT, respectively). Several spike trains, together with their rates estimated by continuous EBM and discontinuous HMM rate estimators, are presented in Fig.~\ref{fig:spiketrainsbio}.

\section{Discussion}

In this paper, we selected alternative coding hypotheses for individual spike trains. For this purpose, we compared rate estimators of the continuous EBM and discontinuous HMM, respectively representing analog and digital neuronal codes. We first determined whether the class of rate process could be identified from synthetic spike trains derived from OUP and SSP. Next, we applied our analytical method to biological data obtained from the visual cortical areas V1 and MT, and the thalamus LGN, and found significant differences among the firing patterns of different brain areas.

Here we assumed two hypotheses; that information transmitted by neurons is coded in an analog or digital manner. If the purpose of selecting coding hypotheses is to best estimate the unknown underlying rate, more coding hypotheses can be accommodated by adopting a suitable model selection principle. For instance, we suggested that two-state HMM represents discontinuous rate processes, but numerous variants are possible. For example, the number of states can exceed two; the number of firing rates is not necessarily fixed in advance but may change arbitrarily in every switching; the firing rate may fluctuate during the inter-transition interval. Such possibilities could be examined using multistate HMMs, the infinite HMM~\cite{infiniteHMM} and the switching state-space model~\cite{ghahramani2000}.

Throughout this study, we have assumed the inhomogeneous Poisson process, in which individual spikes are independently derived from a given rate function of time. However, it should be noted that spiking events are significantly influenced by their predecessors. Consequently, real neuronal firings are not precisely modeled by Poisson processes~\cite{gershon1998,shinomoto2009}. Thus, one may extend our analysis to contend with deviation from Poisson firing, as has been done in Refs~\cite{Tokdar2010,koyama2013}.

Nevertheless, assuming the simple Poisson process is suitable for diverse problems due to its general applicability. In random point processes such as earthquakes, machine noises, and human communications, it would be worthwhile to examine whether the underlying condition is better interpreted as active/inactive, or continuously fluctuating.

\section*{ACKNOWLEDGMENTS}

This study was supported in part by Grants-in-Aid for Scientific Research to SS from the MEXT Japan (25115718, 25240021), and by JST, CREST.


\end{document}